# Criticality in memristor devices and the creation of deep memory

**Stavros G. Stavrinides[1] and Yiannis Contoyiannis[2]**

[1] Physics Department, Democritus University of Thrace, St Lucas Campus, Kavala, Greece
[2] Department of Electrical and Electronics Engineering, University of West Attica, Athens, Greece

**Abstract**: In the present work we describe a way to assess memory capability of real devices, while proposing to the engineering community what to pursue to create devices with *deep* associated memory capability. The study of the signal produced by a real memristor nano-device focused on the description in terms of the Landau $\phi^4$ theory for the critical phenomena in finite systems. This further allowed the utilization of the property of the anomalous enhancement of the autocorrelation function when a system is on the Spontaneous Symmetry Breaking (SSB), for improving the quantity of the demonstrated memory, while simultaneously maintaining a very good quality, as this is expressed by the stability of the autocorrelation function. In this proof-of-concept case, the morphology of the signal allowed us to impose the appropriate modifications on the signal so that we finally show how to get very close to the characteristics of the SSB and thus achieve our goal to get as close as possible to the ideal behavior of a Memristor that yields *deep* memory. Finally, we provide proof of the stability of memristor's operation by showing that solitons "follow" as a skeleton structure the experimentally derived time series.

## 1. Introduction

Memory volatility and the bottleneck between computing and storing resources are two of the main problems faced in computer hardware architectures. The pursuit of ideal memory elements has been attracting the interest of the scientific and the engineering community since the very beginning of the existence of computers. Additionally, a great technological challenge is the creation of elements that are able to process and store information at the same time. Memristors are considered as such promising devices.

Memristors are nonlinear electrical elements, introduced in 1971 by Leon Chua [1], relating the electric charge $q$ with magnetic flux $f_m$. Later, Chua and Kang generalized the concept to memristive systems[2]. The magnitude describing a memristor is memristace, being in fact a resistor demonstrating a memory and it is defined as follows:

$$M(q) = \frac{df_m}{dq} \qquad (1)$$

This definition results to a symmetry between the four (including now memristor) fundamental electrical elements, i.e., resistor, capacitor, inductor, and memristor. Ideal memristors demonstrate some typical features like the current-voltage characteristic pinched hysteresis loop or the degeneration of any hysteresis in the device's resistive behavior to a typical Ohmic behavior when the frequency of the applied signal tends to infinity.

The dynamic behavior of a memristor's resistance, i.e., its memristance, includes the influence of the device's past voltage across or current through it, to its current state. This property allows for the device to demonstrate (at various levels) the so-called non-volatility property [3]. This memory preservation property is ideally expected to hold even when the device is powered off, in the sense that memristance begins from the last value it had just before power interruption [4]. Due to these unique and verypromising features, memristorshave attracted the interest of the scientific and engineering community, both on a theoretical and an experimental level[XX].Thus, a variety of memristor applications have been proposed during the last 15 years, like novel memory nanodevices[5], [6], [17], [18],innovative sensor devices[19], programmable logic [7], control systems [11], reconfigurable computing [12],in-memory computing [13], and Artificial Intelligence [15] (bio-inspired systems like physical neural networks [10, 16].

However, the critical question is if such an ideal or generic memristor device can be realized, with skeptisists claiming that memristor is a purely mathematical concept, [20], [21], [22], [23]. In this work, we investigate the question of how far the mathematical concept (ideal memristor) lies fromreal memristordevices, in terms memory preservation, which is the main characteristic of memristors. This investigation is performed by studying and assessing the current time series coming from a memristor nanodevice. Toward this, we utilize the autocorrelation function and the apparent claim that strong memory preservation is apparently related to and expressed by strong temporal correlations of memristor current (or voltage, for reasons of duality) time series points.

Generally speaking, a system'sor a process's long memory is a property observed and assessed when analyzingaproduced time series. This way, a process or system is considered to demonstrate long memory if the produced time series serial dependence, in other words the autocorrelation function, decays slower than an exponential decay (a time series with an exponentially decaying autocorrelation functionis known as having short memory). This indicates that in long memory time series, the autocorrelation decays hyperbolically and a significant dependence exists between two points even when they are far apart [24].

On the other hand, within the theory of critical phenomena autocorrelation function is considered as a fundamental quantity measuring *memory persistence*of the fluctuations in the order parameter[25]*,* for instance in the case of magnetization in thermal systems.Concepts coming from the theory of critical phenomena, such as self-similarity, power-law distributions, critical exponents, self-organization, spatiotemporal correlations, percolating phase transition, etc., have already been utilized, also in the case of memristors and memristor arrays to simulate the complexity of the brain and to achieve critical dynamics and the consequent maximal computational performance [26].

The motivation for this study, i.e., to investigate memory preservation of a real memristor device in the context of critical phenomena, comes from the following two points:

- According to [27] the drain current in nanoscale fully depleted ultra-thin body and buried oxide n-MOSFET (UTTB-MOSFET) is studied in terms of dynamics of the $\varphi^4$ critical theory (mean field theory [28], [29]). In that work, the measured drain current time series has the form of random telegraph noise (RTN). Thus, in the present work we attempt to extend this study to a memristor nanodevice device, demonstrating RTN in the currentthrough it. This way we will achieve the first goal of our work, which is to interpret the results of this memristor device within the theory of critical phenomena.
- According to [30],the emergence of a very interesting phenomenon within the the $\varphi^4$ critical theory for finite systems, for which we have coined the name"anomalous behavior of the autocorrelation function in the spontaneous symmetry breaking (SSB) hysteresis zone", is described and studied. In the present work,we utilize this phenomenon to achieve a larger relaxation time in the autocorrelation function, something that indicates a longer memory; resulting to a proposal on what should happen to a memristor device to possess a long memory and how to monitor this feature.

Finally, in the present work we go one step further by involving the Kink type solitons [28],

[29] of the $\varphi^4$ critical theory that will guarantee the stability of the signal produced by the memristor that carries the memory information.

## 2. Theoretical background

### 2.1. The second-order phase transition: The 3D-Ising model

Beginning from a generalized spin system $Z(N)$ of $N$ variables, these are defined as:

$$s(a_i) = e^{i2\pi a_i/N} \qquad (2)$$

with $a_i = 0,1,2,3 \ldots N-1$ and $i = 1 \ldots i_{max}$ being the lattice vertices. For $N=2$ the well-known Ising models are produced.

Producing such algorithmic configurations, one may utilize the Metropolis algorithm. In this algorithm the configurations at constant temperatures are configured by Boltzmann statistical weights $e^{-\beta H}$, where H stands for the Hamiltonian of the system's spin. Then the nearest neighbors' interactions can be written as:

$$H = -\sum_{<i,j>} J_{ij} s_i s_j \qquad (3)$$

It is known that such a model undergoes a second-order phase transition when temperature drops below a critical value [25].

For a three-dimensional $20^3$ lattice (3D-Ising model) the critical (or pseudocritical for finite size lattices) temperature has been found to be $T_c = 4.545$ ($J_{ij} =1$) [25]. Considering a sweep of the lattice as the algorithmic time unit and the possible values that spin is allowed to get, being ±1, then the fluctuation of the mean magnetization M is the order parameter.

### 2.2. The second-order phase transition in finite systems

Recently, it has been shown during the second-order phase transition in the case of finite systems and in the absence of any external field, a narrow zone of gradual transition from the critical state to the completion of symmetry breaking exists [25], [26]. In thermal systems such as Ising models, this gradual transition consists of two lobes in the distribution of the values of the order parameter, communicatingonewithanother. As temperature drops, the two lobes gradually separate until complete separation, which indicates the SSB in finite systems. Thus, this transition zone has a temperature range between the critical (pseudocritical temperature for systems of finite size) $T_c$, and the temperature in the SSB, $T_{SSB}$[].

Considering the above, it is apparent that in infinite systems the transition from the critical point to SSB happens in an abrupt way, allowing the unstable critical point to be replaced by two new stable fixed points, in symmetrical positions around it. On the contrary, in finite systems the unstable critical point remains and coexists with the new stable fixed points, as long as the temperature zone exists, i.e., the complete separation of the two lobes.This zone between the critical point and the SSB is a hysteresis zone (with reference to the critical point).

Performing the numerical experiment, using the Metropolis algorithm, for $N_{iter}$=200000 lattice sweeps, one may get the fluctuations of the order parameter in the form of a timeseries. The emerging distributions of mean magnetization Mvalues for three characteristic temperatures appear in Fig. 1. In Fig. 1a,the distribution at the critical point at $T_c$=4.545 is shown. This is the beginning of the hysteresis zone which ends at the conclusion of the SSB at $T_{SSB}$=4.45, appearing in Fig. 1c. In Fig. 1b, the mean magnetization M distribution in the case of an intermediate temperature at $T$=4.49, appears. It is apparent that in finite systems, during the transition, the two lobes communicate with oneanother.

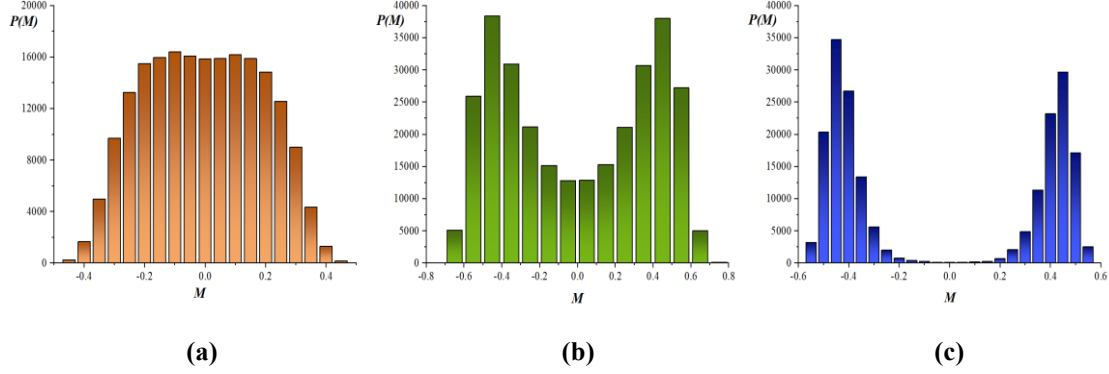

**Fig. 1.** *The distribution of the order parameter (magnetization) in 3D-Ising model for three values of the control parameter (Temperature): (a) Pseudocritical temperature $T_c$=4.545, (b) T=4.49, (c) $T_{SSB}$=4.45.*

The zone's width $\Delta T = T_c - T_{SSB}$ is depended on the length through a power-law of the form $\Delta T \sim d^{-m}$ (m>1) and $d$ the length of Lattice. This means that for smaller systems the width of this zone is larger and apparently, for infinite systems this zone $\Delta T$ vanishes. Therefore, one may consider the established theory for the SSB phenomenon in the infinite limit.

For $N_{iter} \to \infty$, a perfect symmetry between the distribution of negative and positive lobes, for each temperature inside zone, exists. This ensures that the mean value of the magnetization M in the 3D-Ising model remains zero within the zone.

*2.2.1. The dynamics of the critical state*

At the critical state, the dynamics of the fluctuations of the order parameter (M) can be determined analytically for a large class of complex systems by introducing the so-called critical map, which can be approximated as a 1D-intermittent map of type I [32]:

$$\phi_{n+1} = \phi_n + u \cdot \phi_n^z + \varepsilon_n \qquad (4)$$

where $\phi_n$ is the $n^{th}$ sample of the scaled order parameter, $u > 0$ is a coupling parameter, $z$ stands for a characteristic exponent associated with the isothermal exponent $\delta$ as $z = \delta + 1$, and $\varepsilon_n$ stands for the non-universal stochastic noise necessary for the creation of ergodicity.

As is known in the critical state, scaling laws describing the dynamics of the order parameter fluctuations exist. Becoming more specific, by following the Method of Critical Fluctuations (MCF) [31], the laminar length (L) distribution is a power-law of the form [32]:

$$P(L) \sim L^{-p} \qquad (5)$$

Laminar lengths in the case of a time series ($\phi$), correspond to the waiting times, i.e., the number of consecutive time series values that remain within the interval $[\phi_0, \phi_L]$ ("laminar region"), bounded by the fixed-point $\phi_0$ and a number of different values within the $\phi$ values range, which are called "ends of laminar regions" and denoted as $\phi_L$[34], [35](for a detailed description of the MCF analysis methodology with specific examples refer to [31]).

Finally, for the power-law exponent $p$, appearing in Eq. (5), it holds that in the case of critical dynamics [32]:

$$p = \frac{z}{z-1} = 1 + \frac{1}{\delta}, \qquad (6)$$

and in the case of criticality $\delta > 1$, follows that[20]:

$$1 \leq p < 2, \tag{7}$$

Following the above, in the critical state it is expected for finite systems that within the hysteresis zone, where the critical point still exists, the scaling laws will continue to hold exponentially, something that has been proved in [26].

*2.2.2. The phenomenon of anomalous enhancement of correlations inside the hysteresis zone*

In the case of the 3D-Ising model, the order parameter fluctuations demonstrate the On-Off intermittency type, inside the hysteresis zone. In Fig.2 segments of the mean magnetization time series in different situations (in correspondence to the ones in Fig. 1), are presented. In Fig. 2a, the time series corresponding to the distribution of Fig. 1b, for T=4.49 (within the hysteresis zone), is presented, while in Fig. 2b a segment of mean magnetization time series at $T_{SSB}$ (corresponding to Fig. 1c) is shown too.

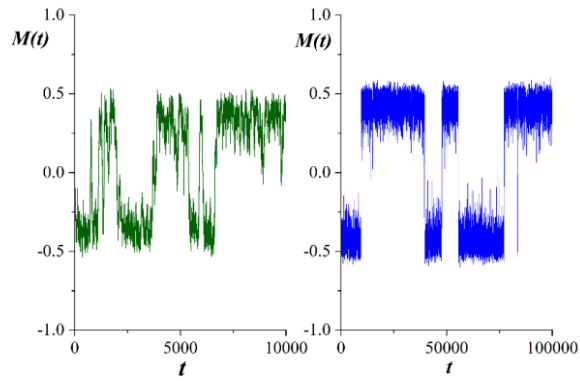

**Fig. 2.** *(a) A segment of the mean magnetization time series M(t) at T=4.49. The resulting behavior is a mixture of Type I intermittency and On-Off intermittency (closer to on-off intermittency). (b) A segment of the time series at $T_{SSB}$. In this case a clear On-Off intermittent behavior is evident. Notice that the time scale in Fig.1b is ten times greater than that in Fig. 1a.*

The emerging dynamics reveal a gradual transition from intermittency of type I to On-Off intermittency [31], where bursts are replaced by jumps between the two levels of the time series values. This means that the system now spends more time in its laminar regions, further meaning that it enhances its capability to retain its memory exactly at the SSB, where bursts are steep, and thus, the waiting times between jumps are maximized. This could be quantitatively expressed by calculating the autocorrelation function $C(m)$, defined for a given time series $x = \{x_1, x_2, x_3, \ldots \ldots x_n\}$ by:

$$C(m) = \frac{\sum_{i=1}^{n-m}(x_i - <x>)(x_{i+m} - <x>)}{\sum_{i=1}^{n}(x_i - <x>)^2} \tag{8}$$

with *m<n* and

$$<x> = \frac{1}{n}\sum_{i=1}^{n} x_i \tag{9}$$

In the case of the intermittent time series of the mean magnetization, the maximization of the laminar lengths would result to the maximization of the autocorrelation function, as well.

This unusual behavior of the autocorrelation function consists of a maximization of its value exactly at the SSB temperature, resulting in getting the highest value than in any other zone temperature. Thus, the autocorrelation function ceases to have its maximum value at the critical point, as the theory of critical phenomena introduces infinite systems. For finite systems, this maximization appears at lower temperatures, exactly where the SSB emerges. In Fig. 3 we present the autocorrelation functions of the 3D Ising model ($m$) for the temperatures appearing in Fig. 1 (with color correspondences).

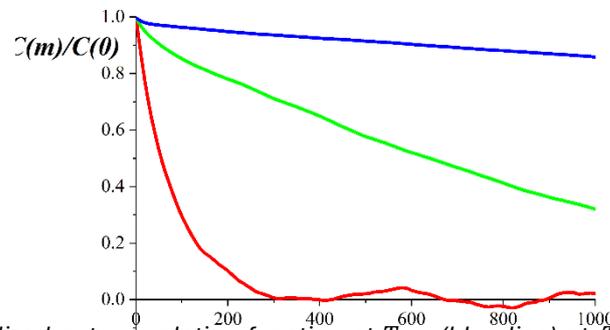

**Fig. 3.** *The normalized autocorrelation function at $T_{SSB}$ (blue line) at $T=4.49$ (green line) and at $T_c$ (red). The phenomenon of the anomalous increase in the autocorrelation function is apparent. Notice that at the end of the hysteresis zone. i.e., the SSB, the maximization of the autocorrelation function takes place.*

*2.2.3. Summarizing the dynamics of second order phase transition*

For a finite system, its dynamics are encompassed on captured time series of any state variable of the system. Considering a second-order phase transition and trying to verify if a system's state is in the critical hysteresis zone, just before the SSB takes place, one may check the following, as these are mentioned in the form of steps to be taken:

*Step 1:* The existence of two lobes in the order parameter distribution that communicate with each other (as in Fig. 1b).

*Step 2:* The emergence of On-Off intermittency dynamics that become predominant as we approach the point of separation of the lobes, i.e. the SSB (as in Fig. 1c).

*Step 3:* The emergence of the phenomenon of the anomalous behavior of the autocorrelation function, while approaching close to the SSB (as Fig. 3 - blue line).

*Step 4:* The laminar length distribution of the order parameter holds a power law with exponent $p$ in the range of values [1,2], for every temperature inside the hysteresis zone.

## 3. Memristor nanodevice and experimental data

In [38], the emerging chaotic dynamics in the case of a memristor nanodevice were experimentally revealed and studied through the experimentally measured Random Telegraph Noise (RTN) coming from properly biased CNM-IMB Ni/HfO2 unipolar Resistive RAM memristor nanodevices. The studied Ni/HfO2-based resistive switching memristor devices were fabricated on (100) n-type CZ silicon wafers, having 7–13mΩ·cm resistivity and a sample appears in Fig. 4. For technical details regarding the manufacturing procedure and a photograph of the actual fabricated device one may see [38].

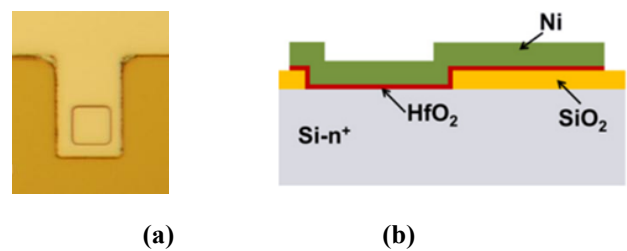

**Fig. 4.** *The unipolar Ni/HfO$_2$ memristor device: (a) the actual fabricated device, and (b) its schematic cross section [38].*

Established tools for confirming and assessing any possible nonlinear behavior, like Grassberger-Procaccia and Cao's method, FNN, Lyapunov exponents as well as surrogate test, were utilized. All these methods and tests clearly suggested the demonstrated RTN time series was not noise (random signal) but a strong complex (chaotic) signal, coming from a deterministic system. It was exactly this deterministic behavior that led us to analyze the time series coming from such devices within the context of the theory of critical phenomena. We remind the reader that such an analysis has been previously performed in the case of the drain current of a in the case of a nano-MOSFET [27]. In Fig. 5 the current, through the studied hereby memristor device, demonstrating the form of a typical RTN is presented, together with the accompanying distribution of the demonstrated laminar lengths.

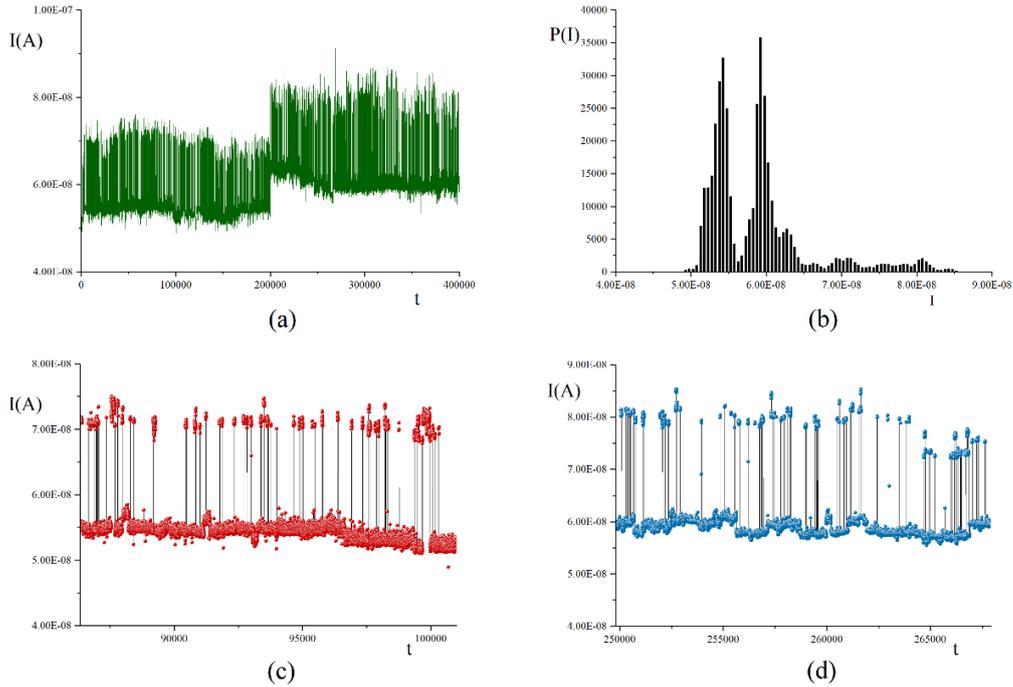

**Fig. 5.** *(a) The RTN current time series of the studied memristornano-device current time-series, (400000 points). (b)The corresponding distribution of the current time series values. The existence of four lobes (two main lobes and two at the tails) is evident. (c) and (d) Detailof the two parts making up the studied timeseries of the Low and High branches correspondingly.*

## 4. Results

As already mentioned, in this work we aim in facing the important property of the memory demonstrated by memristors through the principles posed by phase transitions and the theory of critical phenomena for finite systems. Based on[30] (second order phase transition in finite systems), we will interpret memory production in the framework of the critical phenomena and propose what should happen for the device to demonstrate a "*deeper*" memory.

### *4.1. Criticality in the memristor device*

In the studied hereby case, we define the two phases as the phase that the device generates the RTN current (disorder phase) and the phase of the clearly oriented electron flow (order phase). This definition fully corresponds to the magnetization case, where the phase of high symmetry is expressed by the random orientation of spins, while the phase is the low symmetry phase, with the appearance of domain walls.

The studied time-series is more complicated than the timeseries coming from numerical experiments of the 3D-Ising model. Looking at Fig. 5a, it is apparent that two coarse-graining regions appear. The first part of the timeseries reaches lower values levels (LOW), while the second part reaches higher values levels (HIGH). By zooming intoeach of these regions (for the LOW region look at Fig. 5c and for the HIGH region at Fig. 5d),one may find the existence of On-Off intermittency, as happens in the cases of the hysteresis zone in the Ising model (Fig. 2 step 2), with the differencethat there is an accumulation of points at lower levels.

In Fig. 5b, the distribution of the studied time series clearly shows the appearance of two main lobes ready to separate. At the same time these lobes are followed (for higher values at the tail of the distribution) by a series of smaller lobes.This observation provides us with a hint of the possible existence of SSB (according to step 1).By accepting that the state described by the distribution of Fig. 5b is a coarse-graining approximation of the SSB, we procced to the third step, i.e., examine if the timeseries demonstrates a long temporal memory on very high values, as expected by an ideal memristor. It is expected that the maximization of the auto-correlation function in the SSB occurs when the system gets very close or exactly above the complete separation of the two lobes in the distribution diagrams of the time series values.

Toward this, we calculated the autocorrelation function C(m), according to eqs. (8) and (9), and the results are presented in Fig. 6.Looking at the autocorrelation function C(m) graph in Fig. 6, one may realize that the demonstrated stability of the memory is impressive, since C(m) after 5000 time-stepsremains at an almost constant value, but with the disadvantage of being rather low (<30%).

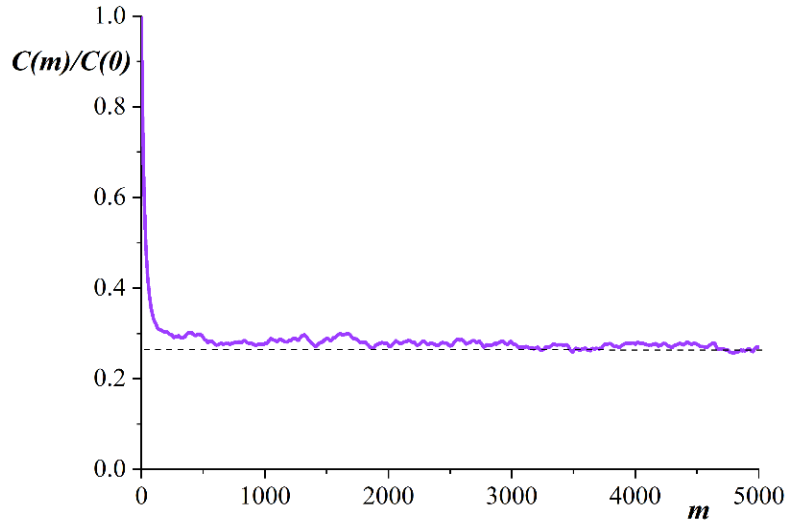

**Fig. 6.** *The normalized auto-correlation function for the memristor time-series in Fig.4a.*

Since we have not yet reached exactly the SSB and attempting to understand the exact state of the system, we utilized the representation of the system in the phase space. In this approach we simply applied the rule of the variable and its numerical derivative (in this case it coincides with the first differences of the timeseries since numerically $\Delta t=1$). The resulting graph appears in Fig. 6b, together with the phase space from the timeseries points of the 3D-ising model for $T_{SSB}$ appearing in Fig. 6a, for comparison reasons.

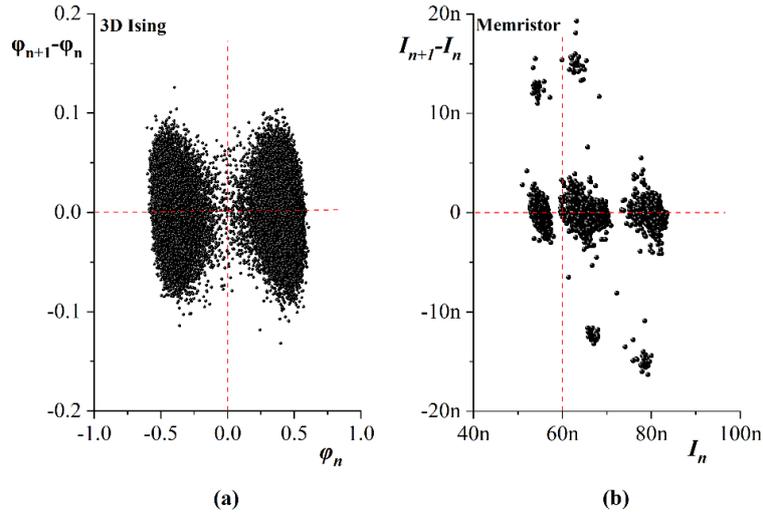

**Fig. 7.** *The phase space for: (a) the 3D-Ising model corresponding to the distribution in Fig. 1c, clearly depicting the separation between the two lobes at the SSB, and (b) the phase space of the experimental time series distribution in Fig. 4b, while the memristor is in a state of RTN. In the latter case several basins of attractors emerge.*

This fragmentation of attractors, in the case of the memristor device, has weakened the correlations, something reflected in the low autocorrelation values appearing in Fig. 6. This confirms the fact that the studied system is far from the ideal SSB that would ensure the phenomenon of the anomalous behavior of the autocorrelation function.

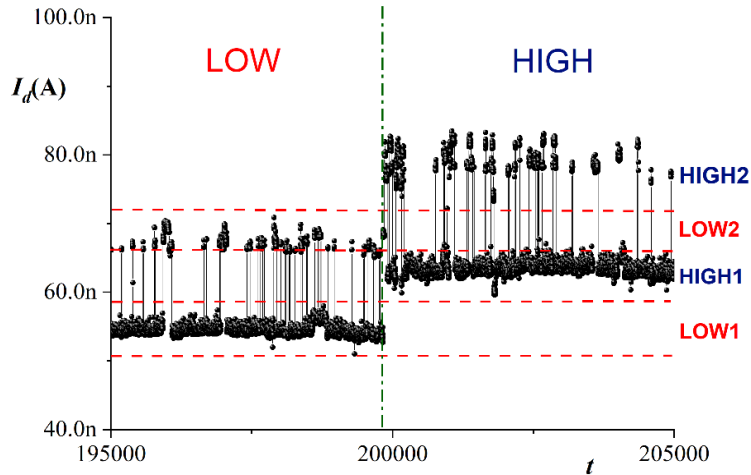

**Fig. 8.** *A detail of the sub-timeseries LOW and HIGH, around transition point from one regio to the other. In this graph, we denote three subregions corresponding to the four lobes appearing in the distribution appearing in Fig. 5b.*

In Fig. 5c and 5d, the two different regions of the studied memristor RTN time series appear, named as the 'LOW' and 'HIGH' sub-time series. Diving into the studied time series properties, we further distinguish four zones, drawing three zone-boundaries (red lines) corresponding to three thresholds, as shown in Fig. 8, where we present a detail of the time series including both regions. In this figure, the LOW1 zone is defined as the lower part of the LOW On-Off intermittency, which corresponds almost to the first main distribution lobe in Fig. 5b. The HIGH1 zone is defined as the lower part of HIGH On-Off intermittency, corresponding almost to the second main lobe in Fig. 5b. Above these zones (LOW1 and HIGH1) the LOW2 and HIGH2 zones appear, each corresponding to the smaller lobesappearing in the tail of distribution.

The autocorrelation function of the time series is mainly determined by the LOW1 and HIGH1 regions,since in these regions concentratemost of the points of the time series, by far. This means that there are long waiting times that keep the autocorrelation function in higher levels as compared to the waiting times appearing in the LOW 2 and HIGH 2 zones. Thus, one may regard the corresponding memory capability of the system as being kept in higher levels, as well.

The existence of the LOW2 and HIGH 2 zones are due to theOn-Off intermittency, which breaks the zones of multiplicity of points (LOW1 and HIGH1), further interrupting the lengths of the waiting times. As a result, the calculated autocorrelation function (Fig. 6) appears to have low values, reducing the system's memory capability.

Considering the above, one may think that if we "cut" the LOW2 and HIGH2 zones, which are distinct and correspond to the very small lobes appearing in the tail of the time series distribution in Fig 5b, the lengths of the waiting times will not only be kept in as high values as possible; further pushing the autocorrelation function to higher values, as well. This would result instrengthening the memory property of the device, moving to what could be called "*a deeper memory property*". Notice that excluding the removal of the LOW2 and HIGH2 regions does not affect the transition region between the two main lobes (see Fig. 8).

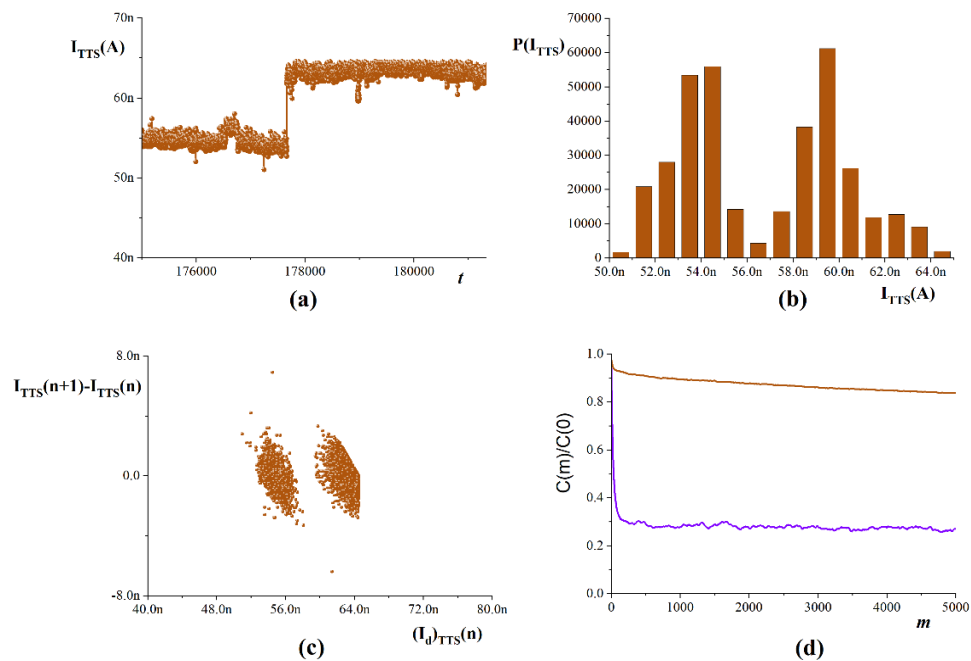

**Fig.9.**(a) A detail of the truncated time series (TTS) around the transition point. (b) The distribution of the values of the TTS; apparently, this state is very close to SSB. (c) The phase space diagram for the TTN confirms the SSB when compared to the corresponding diagram in the 3D-Ising model at the SSB (Fig. 7a). d) The autocorrelation function for a quite large step number (5000). The phenomenon of anomalousenhancement of the autocorrelation function where the autocorrelation is maximized in SSB is obvious.For comparison reasons, we present with the red line the autocorrelation function for the initial time series(Fig. 6).

Thus, it would be interesting to investigate the memory property of the memristor device in the case that no intermittent behavior takes place. Toward this,we created a new time series where the lobes in the tail of the distribution in Fig. 5b were cut-off, i.e.,the values exceeding the threshold of 6.55μA(the red line between the HIGH 1 and the LOW2 zones in Fig. 6). This way the time series gets the form of Fig. 9a. As we see in this figure the truncated time series (TTS) consists of two regions of high multiplicity of points (the LOW 1and HIGH1, with a total of 354 000 points), composing a new On-Off intermittency of a single jump only; of course. This way now there are no "memory gaps" or breaks.

Looking at the distribution of the TTS (Fig. 9b), it is apparent that thisconfirms the existence of a new on-off intermittency of a single jump. The emerging distribution is very close to perfect symmetry, since the probabilities of the two areas of the lobes are 50.2% and 49.8%.In Fig. 9c the corresponding phase space diagram is shown, and one may easily confirm its likelihood to the one In Fig. 8bin the case of a typical 3D Ising model. The autocorrelation function, in the case of the TTS, is expected to get its maximum possible value, something clearly confirmed by the corresponding graph (brown line)in Fig. 9d; notice that for comparison reasons we plotted again the autocorrelation function of the original time series of the memristor nanodevice that appears in Fig. 6. The phenomenon of the anomalous enhancement of the autocorrelation function, i.e., the maximization of the autocorrelation function exactly at the SSB is obvious, further leading to a *deep memory* capability for the system.

Thus, we were able to (a) be consistent with the two-lobe theory predicted in the hysteresis zone of $\varphi^4$ theory for finite size systems, and (b) show the phenomenon of the anomalous enhancementof the autocorrelation function by maximizing its valuesat the SSB. Additionally, we propose to the scientific and engineering community what are the conditions for a memristor device to demonstrate the deep memory property, i.e., long and stable memory.

As already mentioned above (section 2, step 4), the scaling laws should still hold within the hysteresis zone, which means that the distribution of the waiting times should verify a power law exhibiting exponents p∈[1,2]. To calculate this distribution for TTN we have used a method (PNA) recently introduced, which is based on the theory of prime numbers[39]. This method is distinguished by its ability to use the waiting time information from both branches of the time series simultaneously. Due to the very large memory of the system, the waiting times are very long and for this reason the distribution was calculated utilizing a few points. However, this is yet another indication of the existence of the critical hysteresis zone. A presentation of the PNA method as well as its application to the memristor time series, is presented in the appendix. Thus, in the following, we present the emerging results of the application of the PNA on the TTS, without providing any further explanation (provided in the appendix).

In Fig. 10, the calculated scaling of the power law for the laminar length distribution (waiting times) of the TTS, is presented. The exponent of the power law is $p = 1.21 \pm 0.12$. Therefore, both the central value and the extreme values for the calculated exponent clearly lie within the interval [1,2]; thus, declaring the existence of critical dynamics.

Considering eq. (6), one may calculate the value of the isothermal critical exponent $p$ inthe case ofthe 3D Ising universality class [40] by replacing the corresponding value for $\delta = 4.8$. Then, the resulting value is $p = 1.21$, exactly the central value calculated for the TTS. Likewise, by replacing in eq. (6)$\delta = 3$, which is the isothermal critical exponent in the mean field theory universality class [25], one may get $p = 1.33$, exactly the upper extreme value calculated for the critical exponent. Correspondingly, y replacing in eq. (6)$\delta = 15$, which is the isothermal critical exponent of the 2D-Ising universality class [25], one may get $p = 1.07$, very close to the lower extreme value calculated for the critical exponent ($p = 1.09$).

The above significant and initially unexpected results clearly show that the value for the

exponent,regarding the distribution of the waiting times, calculated for the memristor nano-device TTS, in the critical state are absolutely compatible with the three universality classes that demonstratesecond-order phase transition, according to the $\varphi^4$ Landau theory.

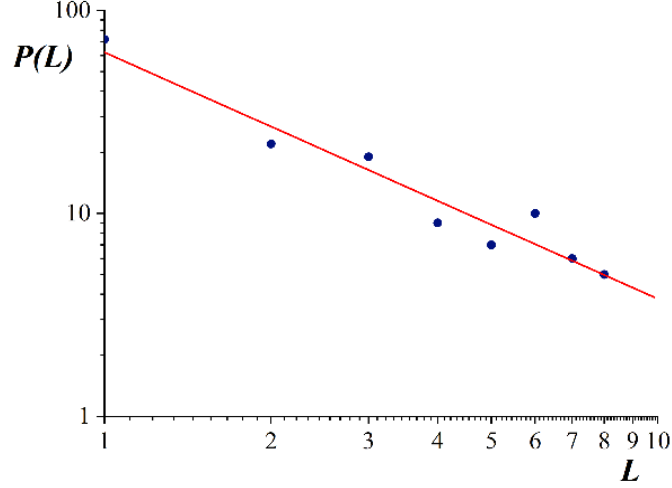

**Fig. 10.**_The distribution of the waiting times of the truncated time series (TTS) of the memristor nanodevice, is presented. Obviously, a power law holds demonstrating an exponent p=1.2 ∈[1,2). Thus, the criterion of criticality is clearly satisfied._

### 6. Kink-antikink $\varphi^4$ solitons

A strong confirmation of all th claims above may come by considering the Kink-type soliton-fields, predicted by the $\varphi^4$ theory for the critical phenomena exactly on the SSB. These fields are symmetric, stable structures describing the transition between the two stable vacua when the SSB takes place. As known, the Landau free energy within the $\varphi^4$ theory is provided by[29]:

$$U(\phi) = \frac{1}{4}(\phi^2 - \frac{m^2}{\lambda})^2 \quad (10)$$

with ϕbeing the order parameter in the second-order phase transition. This free energy demonstrates two degenerate stable vacua. Their values are calculated by [29]:

$$\phi = \pm m/\sqrt{\lambda} \quad (11)$$

And the solitons solution resulting from the above free energy values, are [29]:

$$\phi(x) = \pm \left(\frac{m}{\sqrt{\lambda}}\right) tanh\,[(m/\sqrt{2})(x\text{-}x_o)] \quad (12)$$

with $x$ symbolizing the spatial-temporal space in the classical field theory. These are the Kink and anti-Kink solitons connecting the two stable vacua in SSB.

Having the above in mind, we will attempt to fit the TTSappearing in Fig. 9 by utilizing a fitting function similar to the solution of eq.12, with $x$ as the time component. To facilitate the fitting computational process the y-axis values were multiplied by a factor of 107, while we focused on a small region around the transition point (Fig. 9a) using an arbitrary scale. After arranging the scales, we used as a fitting function the one provided by the following expression:

$$y = p1 * tanh(p2 * (x - p3)\} + p4 \quad (13)$$

Notice that eq. 13 contains soliton's equation (eq. 12), since it is expanded by a constant term necessary for properly displacing the resulting values, since the related symmetry axis should move from zero (as in the case of the soliton within the $\varphi^4$ theory) towards the symmetry axis of the memristor time series, i.e. to the positive values. In Fig. 11 shows the resulting fitting for a segment of the TTS around the transition point is presented. The fitting parameters have the values $p_1 = 0.0497 \pm 0.00017$, $p_2 = 1.43 \pm 0.5$, $p_3 = 660 \pm 0.16$, $p_4 = 0.589 \pm 0.00017$, $R^2=0.98$.

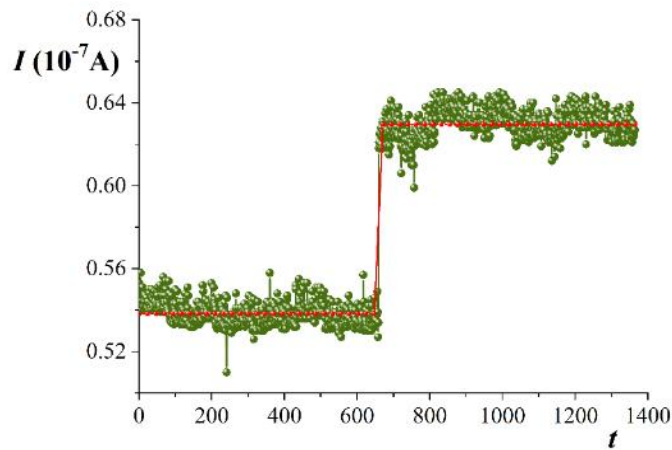

**Fig. 11.** *The fluctuations of the TTN memristor current I for a segment around the transition point appear in green. The red line presents the calculated fitting function of eq.13.*

In this figure one may claim that *"soliton actually meets memristor"*. The fact that a stable structure such as the soliton "follows" as a skeleton structure experimentally derived by a real memristor nano-device TTS, is a proof of the stability of memristor's operation. The existence of fluctuations around the fitting line of soliton, is not only normal but important, as well; and this because the memory capability includes in the TTN is closer to physical reality than the mathematical fitting line.

## 5. MOSFET: The route to SSB

In this section we will make a comparison for memory between the memristor and the MS-DOS which is an analog electronic device that is also in a random telegraph noise state [27]. Figure 12 shows the evolution of the distribution in the hysteresis zone as the control parameter Vg changes for MS-DOS from 300 to 560. As expected, the communication of the lobe's changes.

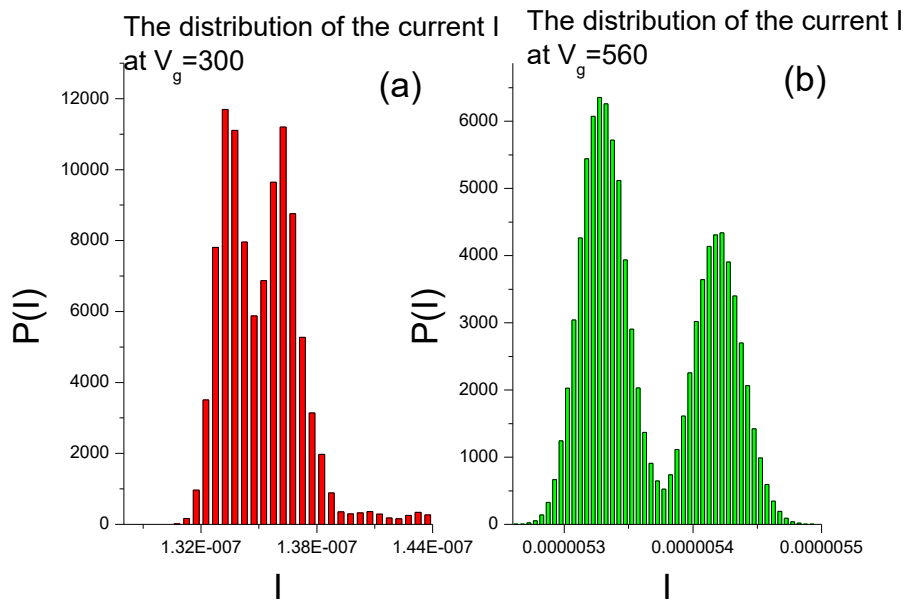

*Fig12. (a) The distribution of the current I at Vg=300 for a MS Dos operated at RTN. (b) The distribution of the current I at Vg=560 for a MS Dos operated at RTN. This distribution is very close to SSB. The fact that one lobe is smaller than the other at 560 is a matter of statistics (100000 points) and is not due to the fact that SSB has already broken since the lobes are still communicating.*

The next fig.13 is present the comparison between the memristor device and the MS-Dos device for their autocorrelations at Vg=300 and Vg=560. So, we use fig 9d where we have the autocorrelation functions for memristor and the ones produced from MS-DOS timeseries which give the distributions of fig 12.

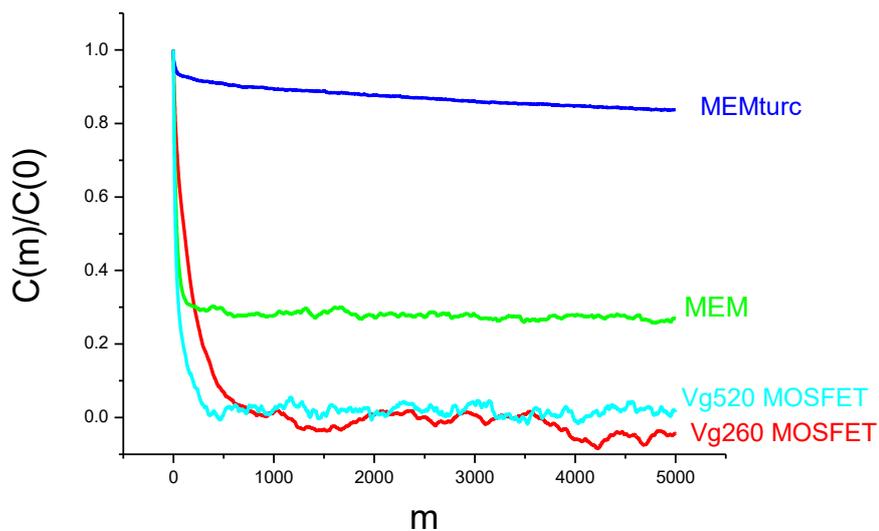

*Fig.13 The normalized autocorrelations for memristor ( (blue, green) and for MS-Dos ( cyan, red) in the same plot for comparison reasons.*

Although they all have the same form , i.e. they are almost parallel to the horizontal axis for large m (m=5000), the difference in the height of the conservation of the correlations value is enormous. Especially if we compare the autocorrelations in the two SSBs, namely the MEMturc (blue) and the MOSFET (cyan) whose distributions have only two lobes without smaller peaks  then we can say with certainty that while the memory of the memristor is huge (between 1 and 0.9), the memory of the Mosfet is at zero. Thus, it appears that for nanodevices that operate in the RTN without an energy-consuming external voltage supply but only with a small bias voltage, the very  high-quality memory is an advantage of memristors . Something like this  is due to the physics of their structure  which favors the existence of almost perfect  criticality in the RTN  state.

## 7.  Conclusions

The main objective of this work is focused on assessing and improving the "quantity" of the memory while simultaneously maintaining the very good quality as this is expressed by the stability of the autocorrelation function.

In the present work we describe a way to assess memory capability of real devices, while proposing to the engineering community what to pursue to create devices with *deep* associated memory capability. The study of the signal produced by a real memristor nano-device focused on the description in terms of the Landau $\phi^4$ theory of the critical phenomena for finite systems. This further allowed the utilization of the property of the anomalous enhancement of the autocorrelation function when a system is on the Spontaneous Symmetry Breaking (SSB), for improving the quantity of the demonstrated memory, while simultaneously maintaining a very good quality, as this is expressed by the stability of the autocorrelation function. In this proof-of-concept case, the morphology of the signal allowed us to impose the appropriate modifications on the signal so that we finally show how to get very close to the characteristics of the SSB and thus achieve our goal to get as close as possible to the ideal behavior of a Memristor that yields *deep* memory.

## Appendix
## A brief presentation of the Prime Number Algorithm (PNA)

The Prime Number Algorithm (PNA), based on the theory of prime numbers, is a method recently published[9], having the unique capability to utilize information from both branches of transition levels, simultaneously. Thus, as shownin [37], [39], it is the most suitable method for determining the distributions of waiting times for time series having a two-branch structure, such as the cases of On-Off intermittency or even for time series with a few transitions between levels.In [39] one may find a detailed presentation of the PNA,as well as the main part of the code of the method. In this appendix, we focus on the application of the PNA tothe time series of symbolic dynamics of two elements that has resulted from the TTSof the memristor current signal, studied hereby.

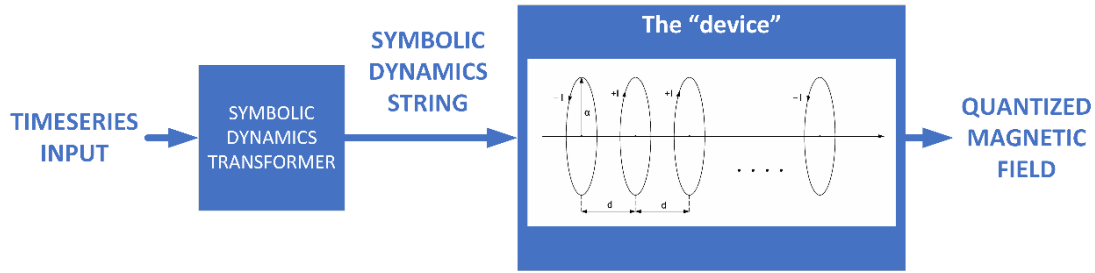

**Fig. 12.** *The general scheme of the PNA method. What is called a "device" is a set of identical rings through which currents of intensity I and random directions flow [41]. The radius of the rings is* a *and the distance of any two consecutive rings is* d.

In Fig. 12, the initial transformation process, which is the heart of the method, is presented, as this is applied in the specific case of the memristor TTS. This initial process includes two stages. In the first stage (left box), a coarse-graining transformation of the timeseries dynamics into a symbolic dynamical string takes place. In this case, all the points with values less than the separation point $5.65 \cdot 10^{-8}$ A(see fig.8b in text) get the value -1, while all the ones having values greater than the separation point get the value +1. In the second stage (right box), the symbolic dynamics string passes through a "device" consisting of identical rings with current flowing through them being clockwise for +1 and counterclockwise for -1. All the rings are having the same radius a and the same distance d between them. As a result, this device corresponds the chronological order of the timeseries points to the positions of the rings in space, since the first incoming time series point defines the value of the current direction at the first ring and so on. The exit of this device is a pattern of quantized values of the total magnetic field, which is produced by applying the PNA on the overall magnetic field at each ring (see details in [39]). In this way a kind o fingerprint of each input is produced; as its structure in the k-position space ("k-waiting times" and gaps) is unique for each time series. In Fig. 13, the output of the device for a small part of the magnetic field (corresponding to a small part of the applied timeseries) is shown; we use only 150 points to be able to show details in the structure of the quantized levels.

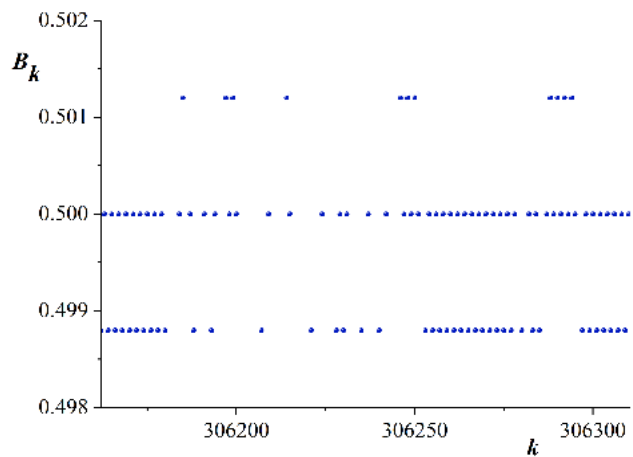

**Fig. 13.** *A segment of 150 points of the magnetic field produced by the "device". The overall pattern has 6 quantized values (3 positive and 3 symmetric negative ones). For the shake of clarity, we*

show only the positive values, in the case that *a=1* and *d=10*, thus *c=a/d = 0.1*. The resulting exactquantized values of the magnetic field (up to four decimal places) are $B_k$=0.5000, 0.4988, 0.5012. Apparently, there are the corresponding negative ones. The lengths of the "k-waiting times" and the gaps that mark these lengths clearly appear.

In [39], it was reported that in the asymptotic limit for which the ratio $c = \frac{a}{d} \ll 1$ only the central levels, i.e., $B_k = \pm 0.5$, remain in the pattern of the quantized magnetic field. Thus, these central values have a special property sincetheir presence does not depend on the value of the geometric characteristics of the device. Consequently, if one wants the calculation of the distribution that takes into account both the positive and the negative subspace of the symbolic dynamics, then the waiting times would be the onescorresponding to the values $B_k = 0.5, B_k = -0.5$, while the rest four values of the magnetic field will interrupt the lengths of the waiting times. This way one may have a quantitative result for the exponent $p$ regardless of the probability of occurrence of the symbols +1, -1.

In Fig .10, the results produced by the PNA according to the above are resulting into a power-law distribution with an exponent 1.21 ∈ [1,2). The comments about this resultscan be followed in the main text.